# Space-coiling Acoustic Metasurface with Independent Modulations of Phase and Amplitude

Pan Li, Qiujiao Du, Meiyu Liu and Pai Peng*

School of Mathematics and Physics, China University of Geosciences, Wuhan 430074, China

Abstract

In this work, we propose a design of acoustic metasurfaces in subwavelength scale enabling independent modulations of phase and amplitude. Each unit cell of the acoustic metasurface consists of simple conventional space-coiling structure added with an air layer, which can be analyzed as two equivalent slabs with non-dispersive effective parameters. The amplitude depends on the space-coiling structure regardless of the air layer, and the phase can be further adjusted by the air layer independent to the amplitude. The acoustic metasurface covers an entire phase change of $2\pi$ and amplitude change of one. We demonstrate an acoustic illusion effect by using the acoustic metasurface screen, which works well as a "full-control" discontinuous boundary to support phase and amplitude differences between the original and illusion patterns. The incident field of a point source is transformed into a target field of the point source scattered by an object with shadow area behind it.

paipeng@cug.edu.cn

In the past decade, the emerging of metasurfaces [1,2], a kind of sheet material with subwavelength artificial structures, make it possible to modulate electromagnetic wavefront. Every unit cell with inertial micro-structures of a metasurface could provide phase delays along the surface, and it opens a new degree of freedom in phase modulation [1]. This concept of metasurfaces are extended to acoustic waves, called acoustic metasurfaces (AMs) [3-17]. Coiled channels [5-8] or Helmhertz resonators [9-11] as micro-structures are drilled on the surface of rigid half-space in order to manipulate reflected acoustic wavefront. Later, rigid slabs opened with those micro-structures are proposed, regarded as transmitted-AMs to control transmitted waves. Due to the large acoustic impedance mismatch between the transmitted-AMs and air background, the transmission coefficient would be generally low [12,13]. Many efforts are made on achieving high-transmission AMs via reducing the effective impedance of AMs (wide [14] or smooth [15] the channels) or borrowing the Fabry–Pérot (FP) resonance [16]. AMs with uniform high-transmission are achieved successfully and widely used in bending or focusing acoustic waves. Although the amplitude can be low or high by adjusting the impedance, to obtain an AM with simultaneous phase and amplitude modulations is still a challenging subject. For some applications with complex patterns, such as the acoustic holography [18-21] or illusion [17], AMs with controllable transmissions are preferred to achieve high-quality patterns.

Very recently, Ye tian et al. [19] firstly proposed a "full-control" AM (FAM) with a decoupled modulation of phase and amplitude based on coating unit cells and perforated panels. Later, FAMs are realized in three-dimensional experiments [20]. These FAMs are not designed under a guidance of theory model. Reza et al. firstly proposed a theory model [21], which theoretically analyze the limitations of conventional AMs (CAM) in achieving simultaneous phase and amplitude modulations. They also design a kind of horn-like FAMs. But the phase and amplitude cannot be adjusted independently. In particular, low amplitudes is still hardly obtained at the FP resonances, where the horn-like micro-structure would be very complex. In these designs [19-21], the internal micro-structures have to increase in complexity, as a cost of the additional capability to manipulate amplitudes. In this letter, we propose a new design of FAM with simple space-coiling structure the same to that in CAMs [7,13-15], enabling simultaneous phase and amplitude modulations independent. The proposed FAM could cover an entire phase change of $2\pi$ and amplitude change of one simultaneously. We demonstrate that the FAM works well in the acoustic illusion effect as a "full-control" discontinuous boundary to support phase and amplitude differences between the original and illusion patterns.

To achieve a FAM, we firstly briefly investigate the limitations of CAMs. We consider a CAM based on the popular space-coiling structure, as shown in Fig. 1(c) topological schematic illustration. The CAM consists of a rigid slab opened with periodic curved slits [13]. The parameters $p=5\lambda/34$, $L=0.75\lambda$, $d=\lambda/136$ and $l=1.5\lambda$ are the periodic constant, length of the unit cell, width and effective total path of the curved slit, respectively, and $\lambda$ is the working wavelength in air. We use the finite element method (with COMSOL software) to calculate the amplitudes $A$ (normalized by incident amplitude $A_0$) and the phases $\varphi$ (normalized by $\pi$) for the transmitted waves as a function of wavelength $\lambda'$ (normalized by $\lambda$) and plot the results in Figs. 1(a) and (b) red lines, respectively. The green lines in Figs. 1(a) and (b) are the amplitudes and phases of a CAM with wide slits (5 times), resulting in low impedance. By comparing the red and green line in Fig. 1(a) we can see that the amplitude is affected by the impedance for most frequencies. In general, high amplitudes can be obtained via changing the impedance, but low amplitudes is hardly obtained near the FP resonances, where the phases equal to $m\pi$. In physics, the FP resonances always lead to full transmission regardless of the impedance, bringing a constraint of amplitudes and phases. That is the constant amplitude $A=1$ bounded with constants phases $\varphi = m\pi$ at the FP resonances. We call the constraint at the FP resonances as the "FP binding", as shown in Figs. 1 blue dashed lines. The limitations of CAMs can be mainly attributed to the FP resonance, where only one option is available for the combination of phase and amplitude.

The FP binding can be easily analyzed when the CAM is regarded as a homogenous equivalent slab, whose non-dispersive effective parameters (such as relative impedance $Z_{eff} = p/d$ and index $n_{eff} = l/L$) can be extracted by comparing transmission and reflection spectra [13]. The analytical expressions of phase and amplitude for transmitted waves of the equivalent slab are, respectively,

$$\varphi = m\pi + \arctan\left[Q\tan\left(k_{eff}L\right)\right] \tag{1}$$

and

$$|A| = \frac{1}{\sqrt{\cos^2\left(k_{eff}L\right) + Q^2 \sin^2\left(k_{eff}L\right)}} \tag{2}.$$

Here $Q = \left(1 + Z_{eff}^2\right)/2Z_{eff}$ is an impedance factor, and $k_{eff}$ is the effective wave vector in the equivalent slab. In case of FP resonances, when $k_{eff}L = m\pi$ is satisfied, the Eqs. (1) and (2) will reduce to $\varphi = m\pi$ and $A = 1$, respectively, showing the FP binding. Equations (1) and (2) could further give the possible range of values for combinations of phases and amplitudes [21] as shown in Fig. 1(c) gray region. The rest white region denotes the forbidden amplitudes, and our goal is to fill up the forbidden range. Although there are certain constraints elsewhere, we can focus on the FP binding for simply. Once if we find a way to break the FP binding, manipulating phases and amplitudes individually at FP resonances, the white region will be somewhat filled and achieving FAM will be available.

On purpose to break the FP binding, complex internal micro-structures are involved, making the AMs no longer regarded as homogeneous slabs in pervious works [19-21]. Here we show that the FP binding is broken by simply compressing the length of the unit cell. As shown in Figs. 2(a) and (b) schematic sketches, we keep the internal micro-structure parameters ($d$, $p$ and $l$) of unit cells but compress the length from $L$ to $h$. The extra space (with length $s = L - h$) after the compression is filled up with air background. The whole AM (remained with length $L$) can be regarded as two equivalent slabs: a compressed CAM slab (with length $h$) added with an air layer (with length $s$). This system is easily analyzed. The analytical expressions of phase and amplitude for transmitted waves can be expressed as:

$$\varphi_F = m\pi + \arctan\left[\frac{Q\tan\left(k_{eff}^F h\right) + \tan(ks)}{1 - Q\tan\left(k_{eff}^F h\right)\tan(ks)}\right] \tag{3}$$

and

$$|A_F| = \frac{1}{\sqrt{\cos^2\left(k_{eff}^F h\right) + Q^2 \sin^2\left(k_{eff}^F h\right)}} \tag{4},$$

respectively, with $k$ the wave vector in air background. Because the total path $l$ is kept before and after the compression, we have $k_{eff}^F h = k_{eff} L = kl$, and Eq. (4) is the same to Eq. (2). That means the amplitude $A_F$ is not affected by the air layer and only depends on the CAM slab. Once the CAM slab is given, the amplitude $A_F$ is determined regardless of the added air layer. In addition, if we apply $Q\tan\left(k_{eff}^F h\right) = Q\tan\left(k_{eff} L\right) = \tan\varphi$ obtained from Eq. (1), the Eq. (3) will reduce to $\varphi_F = \varphi + ks$. It shows that the phase $\varphi_F$ can be a function of the air layer length $s$ independent to the amplitude $A_F$. It is interesting that the two slabs, the CAM slab and the added air layer, can individually play a role to control the amplitude $A_F$ and the phase $\varphi_F$, respectively. In general, any required amplitude $A_F$ can be obtained by adjusting the slit width $d$ and length $l$ inside the CAM slab [13], and any required phase $\varphi_F$ can be obtained by further adjusting the air layer length $s$.

Now we focus on the FP binding. When $k_{eff}^F h = m\pi$ is satisfied for FP resonances, we will have $l = m\lambda/2$ by applying $k_{eff}^F h = kl$ into Eq. (4). Because the total path $l$ is kept, the FP resonance is also remained at the same frequency. Equations (3) and (4) reduce to $\varphi_F = m\pi + ks$ and $A_F = 1$ at the FP resonance, respectively. We can see that the phase $\varphi_F$ are no longer constants but as a function of $s$, and the amplitude $A_F = 1$ can combine with phases other than $m\pi$. On the other hand, when the phase is $\varphi_F = m\pi$, the amplitude $A_F$ will no longer equal to one, and the phases $\varphi_F = m\pi$ can combine with amplitudes other than one.

To clearly show the breaking of FP binding, we choose three typical AMs constructed by different unit cells and plot the transmitted pressure fields, as shown in Figs. 2(a)-(c). Their combinations of phases (reduced in $0 \sim 2\pi$) and amplitudes are $\varphi = \pi$ and $A = 1$ for Case1, $\varphi_F \approx 1.35\pi$ and $A_F = 1$ for Case2, and $\varphi_F = \pi$ and $A_F \approx 0.5$ for Case3, respectively.

In Case1, the structure is the CAM discussed in Fig. 1. The pressure field shows a full transmission at the FP resonance. The gray region in Fig. 2 (d) above the black solid line is the same to the gray region in Fig. 1(c), showing the theoretically possible range for combinations of phases and amplitudes. The blue dot marks the combination for Case1 in the FP binding as a reference.

In Case2, the parameters $d$, $p$ and $l$ are the same to those in Case1, but the length is compressed from $L$ to $h \approx 0.57\lambda$. The full transmission is also observed because the total path

$l$ has not changed and the FP resonance is remained. The transmitted wavefront has a phase shifting $\Delta\varphi \approx 0.35\pi$ comparing to Case1. Because the phase at the output surface of the unit cell has increased by $\Delta\varphi = ks$ comparing to Case1 due to the additional air layer. As a result, although Case2 is still at FP resonance, the amplitude of $A_F = 1$ can independently combine with phases $\varphi_F = m\pi + ks$. The corresponded possible range for Case2 obtained from Eqs. (3) and (4) is shown in Fig. 2 (e) gray region, which has the same geometry to that of Case1, but the region shift $\Delta\varphi = ks$ to the right side, as shown in Fig. 2(e) blue arrows.

In Case3, the parameters $d$, $p$ and $h$ are the same to those in Case2, but the total path $l$ is shorten in order to offset the phase increasing in the air layer. The phase difference of the whole unit cell is kept as $\varphi_F = \pi$ the same to Case1, so the wavefront is also the same to Case1, as shown in Fig. 2(c). But it is on longer full transmission ($A_F < 1$), because the total path inside the CAM slab layer is changed to $l < m\lambda/2$, leading deviating from the FP resonance. The possible range of combinations for Case3 is the same to Case2 because of their same lengths, resulting in the same phase shifting $\Delta\varphi$. The solid red dot in Fig. 2 (e) marks the combination for Case3, whose amplitude shifts down by $\Delta A \approx 0.5$ compared to Case1, as shown in Fig. 2(e) red arrow. Here the amplitude $A_F = 1 - \Delta A$ for Case3 is no longer constant but can change in a range between $A = 1$ and $A_F$. The minimum value of the $A_F$ can be analyzed from Eqs. (3) and (4) as $A_F^{\min} = 1/\sqrt{1 + \tan^2(ks)}$, depending on the phase shifting. The exact value of amplitude $\Delta A$ is effected by the impedance factor $Q$. The Case3 shows that the phase of $\varphi_F = \pi$ can independently combine with other amplitude $A_F$ by changing the air layer length $s$ and the impedance factor $Q$.

In particular, when we break the FP binding, the possible range for combinations of phases and amplitudes will expand due to the additional choices after the breaking. For example, if the length of the CAM slab has two options of $L$ and $h = 0.57\lambda$ as discussed in Case1 and Case2, the possible range will enlarge as shown in Fig. 2(f) gray region, which is merged from the gray regions in Figs. 2(d) and (e). Any combination of phases and amplitudes in Fig. 2(f) gray region is available if we can choose the length to be $L$ or $h = 0.57\lambda$. By extension, if the length $h$ can be any value between $L$ and $h_{\min} = L - \lambda/2$, where the phase shifting could gradually increase from zero to $\Delta\varphi = \pi$, the merged possible range will cover the entire range of values, including the phase change of $2\pi$ and amplitude change of one. That means arbitrary combination of phases and amplitudes is available.

Based on that theory, we build a FAM, as shown in Fig. 3(a), by assembling the compressed unit cells as playing Lego blocks. The unit cells of the FAM have simple space-coiling structures with different lengths. The unit cells are output-justified, and the input surface is uneven. We demonstrate an acoustic illusion effect by using the FAM as an instance for applications. Unlike the pervious illusion devices based on the transformation approach [22,23], here the acoustic illusion effect is achieved by supporting phase and amplitude discontinuities between the original and illusion fields.

As shown in Fig. 4(a), the target illusion field is a pressure field of a point source scattered by a rigid square. The acoustic illusion effect is shown in Fig. 4(b), where a FAM screen is placed behind the point source and the same "scattered" field is obtained. The phase and amplitude profiles of pressure for the point source at the input boundary ($y = 0$) of the FAM screen can be analyzed as a function of positions $\varphi_{\text{in}}(x)$ and $A_{\text{in}}(x)$, respectively. Similarly, the phase and amplitude profiles for the target field at the output boundary ($y = L$) of the FAM screen are extracted from Fig. 4(a) as $\varphi_{out}(x)$ and $A_{out}(x)$, respectively. The phase and amplitude discontinuities are defined as $\varphi_D(x) = \varphi_{out}(x) - \varphi_{in}(x)$ and $A_D(x) = A_{out}(x)/A_{in}(x)$, which are shown in Figs. 3(b) and (c) black circles, respectively. Both the phase and amplitude discontinuities can be obtained by the particular FAM design. As shown in Figs. 3(b) and (c) red squares, the designed phase and amplitude profiles of the FAM are perfectly matched with the required profiles (black circles). As a result, the FAM screen works well as a full-control discontinuous boundary, turning the original incident field into the target scattered field.

As a comparison reference, we also calculate the illusion effect obtained by a CAM with only phase modulations. The phase and amplitude profiles provided by the CAM are shown in Figs. 3(b) and (c) green triangles, respectively. The phase profile is perfectly matched, but the amplitudes are uniform. Large differences between the illusion and target patterns are observed, as shown in Fig. 4(c). The chosen target field features a large shadow area behind the object, which obviously cannot be represented by the CAMs with uniform high amplitudes. Low amplitudes combined with arbitrary phases are hardly obtained by the CAMs. Although some acoustic illusion effects can be achieved by CAMs [17], these effects would limit in simple illusion patterns, such as a source shifter. The illusion error caused by a uniform amplitude distribution will become quite prominent when the target profile is complicated and comprises uneven amplitude levels [20].

In conclusion, we propose a method to build FAM enabling simultaneous phase and amplitude modulations. The unit cells have subwavelength space-coiling structures with different lengths, and the FAM can be easily analyzed because of its simple structure. The FAM could cover an entire phase change of $2\pi$ and amplitude change of one simultaneously. In addition, the phase and amplitude can be adjusted independently. We further demonstrate an acoustic illusion effect by utilizing this FAM, which works well as a full-control discontinuous boundary to support phase and amplitude differences between the original and illusion patterns. The field of a point source is transformed to a scattered field with shadow areas behind an object. This illusion effect with complex scattered field cannot be represent by using the CAMs with phase modulation only.

This work was supported by the National Natural Science Foundation of China (Grant No: 11604307), the CUG Baseline Research Fund (Grant No: G1323521779), the Key Program of National Natural Science Foundation of China (Grant No: 41830537), and National Programme on Global Change and Air-Sea Interaction (GASI-GEOGE-02)

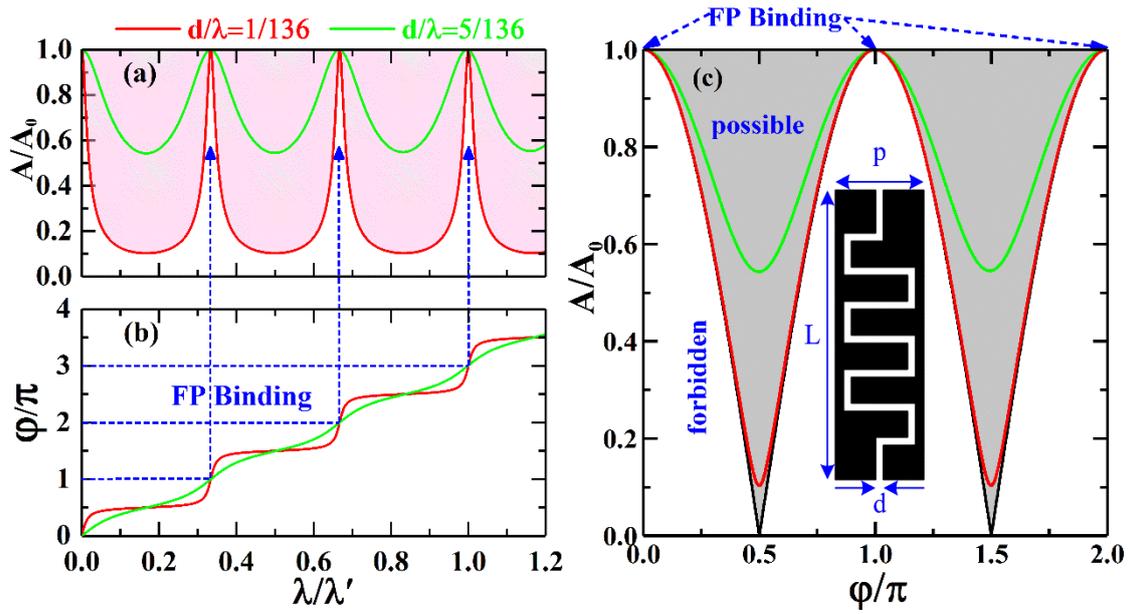

FIG. 1  Red lines are the (a) amplitude and (b) phase spectrum of a CAM with space-coiling structure. Green lines correspond to another CAM with wide slits. Light red region denotes the feasible amplitudes obtained by reducing the impedance. (c) The reduced phase and amplitude range of values. The gray region is the theoretical possible range for combinations of phase and amplitudes, whereas the rest (white) region denotes the forbidden range. Red and green solid lines in (c) corresponds the results in (a) and (b). The illustration in (c) presents the topological schematic geometry of the unit cell.

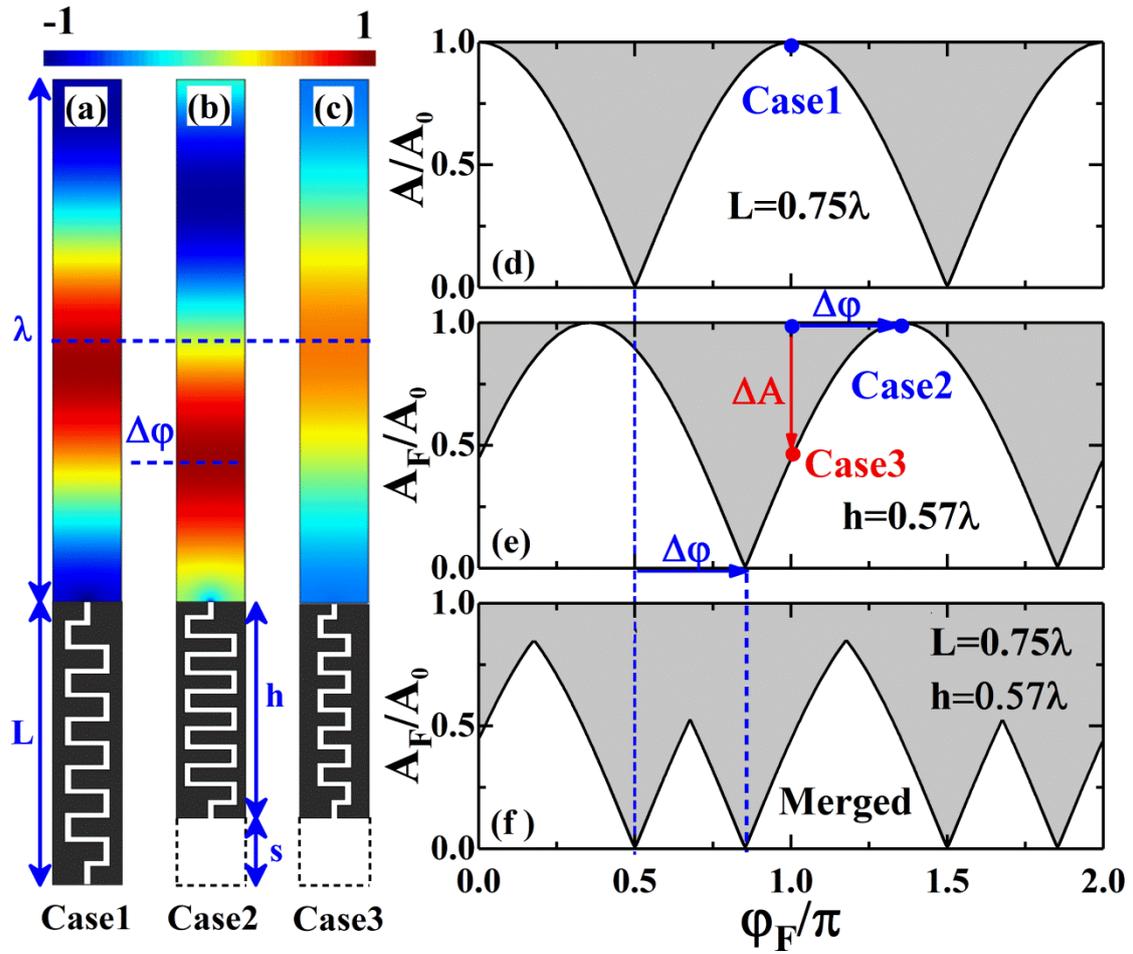

FIG. 2 (a), (b) and (c) are the pressure fields for different unit cells. Blue and red color denotes the minimum and maximum value of pressure. (d) and (e) is the phase and amplitude range of values corresponding to the structure in (a) and (b), respectively. The gray region denotes the theoretical possible range for combinations of phase and amplitudes. Blue and red dots mark the positions for the three cases. (f) is the merged region of (d) and (e).

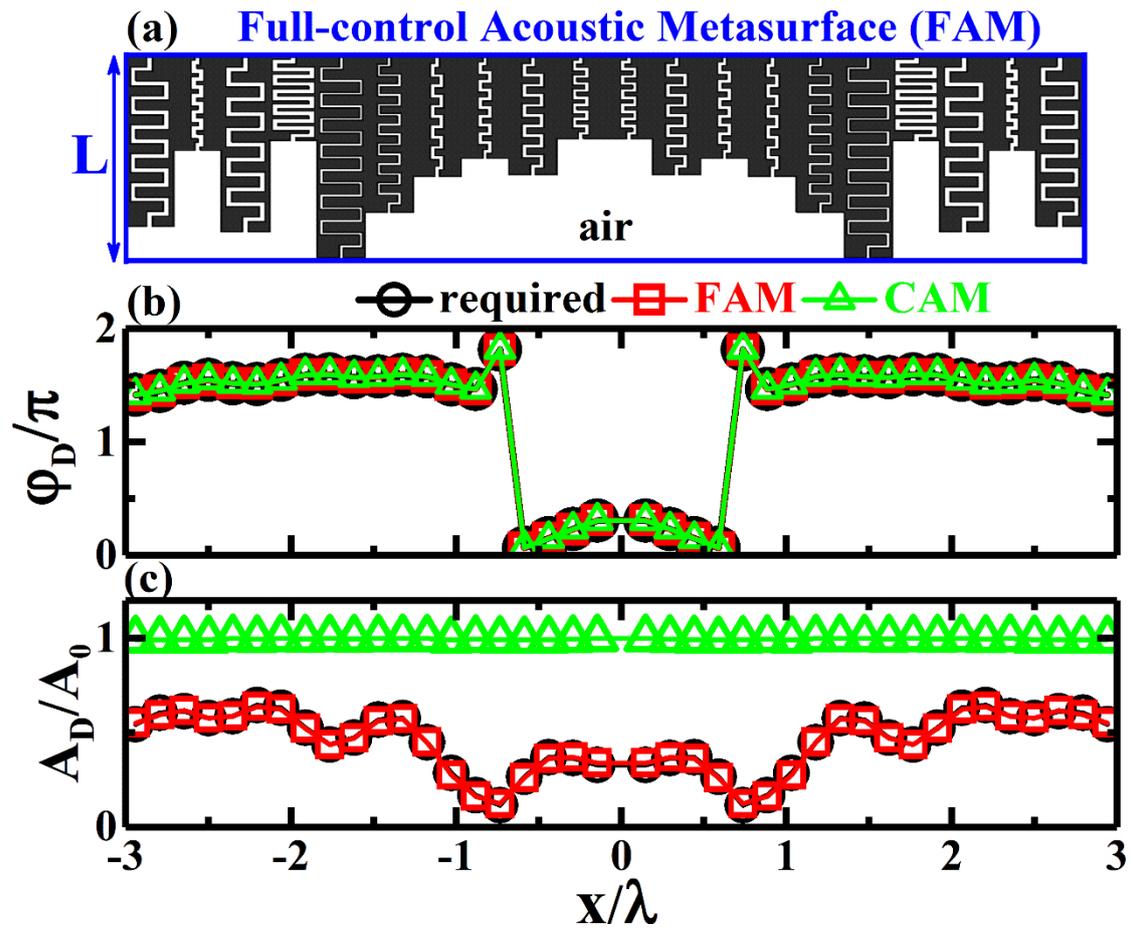

FIG. 3  (a) schematic sketches of a FAM by assembling the compressed unit cells with different length.   (b) and (c) are the phase and amplitude discontinuities profiles, respectively.

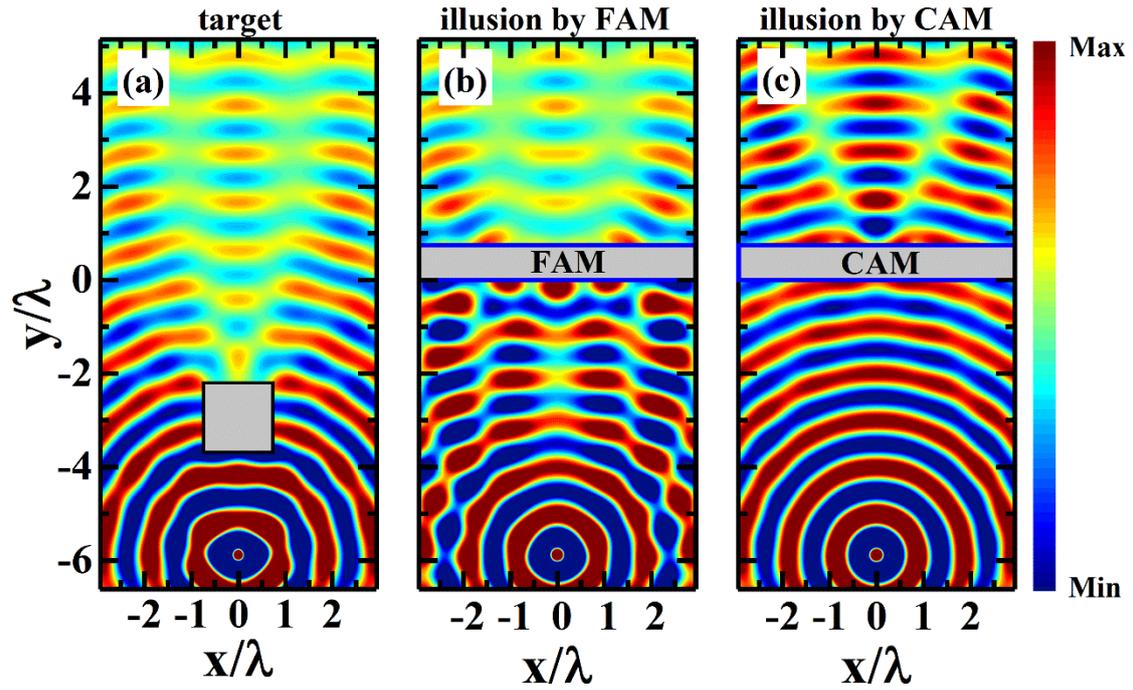

FIG. 4  (a) is the pressure field of a point source scattered by a rigid square as the target field. Blue and red color denotes the minimum and maximum value of pressure.

(b) and (c) are the illusion field obtained by FAM screen and CAM screen, respectively. The point source and the rigid square are placed at $y \approx -6\lambda$ and $y \approx -3\lambda$, respectively, and the length of the square is $a = 1.5\lambda$.

Reference

[1] N. Yu, P. Genevet, M. A. Kats, F. Aieta, J.-P. Tetienne, F. Capasso, and Z. Gaburro, Science **334**, 333 (2011).

[2] A. V. Kildishev, A. Boltasseva, and V. M. Shalaev, Science **339**, 1232009 (2013).

[3] B. Assouar, B. Liang, Y. Wu, Y. Li, J.-C. Cheng, and Y. Jing, Nature Reviews Materials **3**, 460 (2018).

[4] B. Cheng, H. Hou, and N. Gao, Modern Physics Letters B **32**, 1850276 (2018).

[5] T. Weipeng and R. Chunyu, Journal of Physics D: Applied Physics **50**, 105102 (2017).

[6] S. A. Cummer, J. Christensen, and A. Alù, Nature Reviews Materials **1**, 16001 (2016).

[7] Y. Li, B. Liang, Z.-m. Gu, X.-y. Zou, and J.-c. Cheng, Scientific Reports **3**, 2546 (2013).

[8] Y. Li, X. Jiang, R.-q. Li, B. Liang, X.-y. Zou, L.-l. Yin, and J.-c. Cheng, Physical Review Applied **2**, 064002 (2014).

[9] B. Liu, W. Zhao, and Y. Jiang, Scientific Reports **6**, 38314 (2016).

[10] C. Ding, X. Zhao, H. Chen, S. Zhai, and F. Shen, Applied Physics A **120**, 487 (2015).

[11] J. Guo, X. Zhang, Y. Fang, and R. Fattah, Journal of Applied Physics **124**, 104902 (2018).

[12] Y. Li, B. Liang, X. Tao, X.-f. Zhu, X.-y. Zou, and J.-c. Cheng, Applied Physics Letters **101**, 233508 (2012).

[13] P. Peng, B. Xiao, and Y. Wu, Physics Letters A **378**, 3389 (2014).

[14] K. Tang, C. Qiu, M. Ke, J. Lu, Y. Ye, and Z. Liu, Scientific Reports **4**, 6517 (2014).

[15] Y. Xie, W. Wang, H. Chen, A. Konneker, B.-I. Popa, and S. A. Cummer, Nature Communications **5**, 5553 (2014).

[16] Y. Li, X. Jiang, B. Liang, J.-c. Cheng, and L. Zhang, Physical Review Applied **4**, 024003 (2015).

[17] Y. Liu, Z. Liang, F. Liu, O. Diba, A. Lamb, and J. Li, Physical Review Letters **119**, 034301 (2017).

[18] Q. Wang *et al.*, Scientific Reports **6**, 32867 (2016).

[19] Y. Tian *et al.*, Applied Physics Letters **110**, 191901 (2017).

[20] Y. Zhu, J. Hu, X. Fan, J. Yang, B. Liang, X. Zhu, and J. Cheng, Nature Communications **9**, 1632 (2018).

[21] R. Ghaffarivardavagh, J. Nikolajczyk, R. Glynn Holt, S. Anderson, and X. Zhang, Nature Communications **9**, 1349 (2018).

[22] B.-I. Popa and S. A. Cummer, Nature Communications **5**, 3398 (2014).

[23] J. Li and J. B. Pendry, Physical Review Letters **101**, 203901 (2008).